\documentclass[lettersize,journal]{IEEEtran}
\usepackage{amsmath,amsfonts}
\usepackage{algorithm}
\usepackage{array}
\usepackage[caption=false,font=normalsize,labelfont=sf,textfont=sf]{subfig}
\usepackage{textcomp}
\usepackage{stfloats}
\usepackage{url}
\usepackage{verbatim}
\usepackage{graphicx}

\usepackage{svg}
\usepackage{enumitem}
\usepackage{soul}

\usepackage{algpseudocode}
\usepackage[style=ieee,backend=biber]{biblatex}

\usepackage{eurosym}
\usepackage{amssymb} 
\usepackage[normalem]{ulem}
\usepackage{booktabs}

\usepackage{multirow}
\usepackage{makecell}
\usepackage{bm}
\usepackage[table]{xcolor}

\definecolor{orange1}{RGB}{255,240,210} 
\definecolor{DarkOrange}{RGB}{200,90,0}
\definecolor{orange2}{RGB}{255,150,70} 
\definecolor{customgreen}{HTML}{8DCB20}
\definecolor{customlight}{HTML}{D0CCBE}
\definecolor{customgray}{HTML}{D9D9C1}
\definecolor{customwhite}{HTML}{FFFFFF}
\definecolor{valgreen}{HTML}{8DCB20}
\definecolor{relgreen}{HTML}{778A4F}
\usepackage[colorlinks=true,
            linkcolor=black,
            urlcolor=customgreen,
            citecolor=customgreen]{hyperref}
\usepackage{cleveref}  
\begin{document}

\title{PriceFM: Foundation Model for Probabilistic Electricity  \\ Price Forecasting}

\author{Runyao Yu$^{1,2,3}$, 
Chenhui Gu$^1$, 
Jochen Stiasny$^1$,  
Qingsong Wen$^{4,5}$,  
Wasim Sarwar Dilov$^3$, \\
Lianlian Qi$^{2,6}$,  
Jochen L. Cremer$^{1,2}$
\\ $^1$Delft University of Technology, $^2$Austrian Institute of Technology,  $^3$Rimac Technology, \\    $^4$University of Oxford, $^5$Squirrel AI,  $^6$Technical University of Munich}



\maketitle
\begin{abstract}

Electricity price forecasting in Europe presents unique challenges due to \textcolor{black}{increasing renewable generation variability,} market integration, and the continent's physically interconnected power system.
While recent advances in foundation models have led to substantial improvements in general time series forecasting, most existing approaches do not incorporate prior graph knowledge from the transmission topology, which can limit their ability to exploit meaningful cross-region dependencies in interconnected power systems, motivating a domain-specific foundation model.
In this paper, we address this gap by first introducing a comprehensive and up-to-date dataset across 24 European countries (38 regions), spanning from 2022-01-01 to 2026-01-01.
Building on this groundwork, we propose {PriceFM}, a  probabilistic foundation model pretrained on this large dataset. 
Specifically, {PriceFM} maps each region’s price and exogenous features\textcolor{black}{, including load, solar, and wind generation forecasts,} into a comparable latent embedding via a shared Mixture-of-Experts (MoE) projection layer, then injects prior graph knowledge by constructing a sparse graph mask derived from transmission topology. 
Across a large-scale European benchmark, PriceFM achieves strong performance and demonstrates superior generalization compared with multiple competitive baselines. \textcolor{black}{The results highlight the value of topology-guided forecasting with increasing renewable generation and strong cross-border interconnections.}
The methodology is available at: \url{https://runyao-yu.github.io/PriceFM/}.
\end{abstract}

\begin{IEEEkeywords}
\textcolor{black}{Electricity Price Forecasting, Renewable Energy Integration, Probabilistic Forecasting, Transmission Topology, Foundation Models, Sustainable Power Systems}
\end{IEEEkeywords}

\section{Introduction}
\label{sec:intro}

\textcolor{black}{Accurate probabilistic electricity price forecasting is increasingly important for sustainable power-system operation under high renewable penetration~\cite{10461088, 7533504}. Solar and wind generation introduce strong temporal variability and regional heterogeneity, which affect market prices through both local supply-demand conditions and cross-border power exchanges~\cite{lago2018forecasting}.}
However, physical constraints, such as limited transmission capacity, 
\begin{figure}[!ht]
\includegraphics[width=0.97\linewidth]{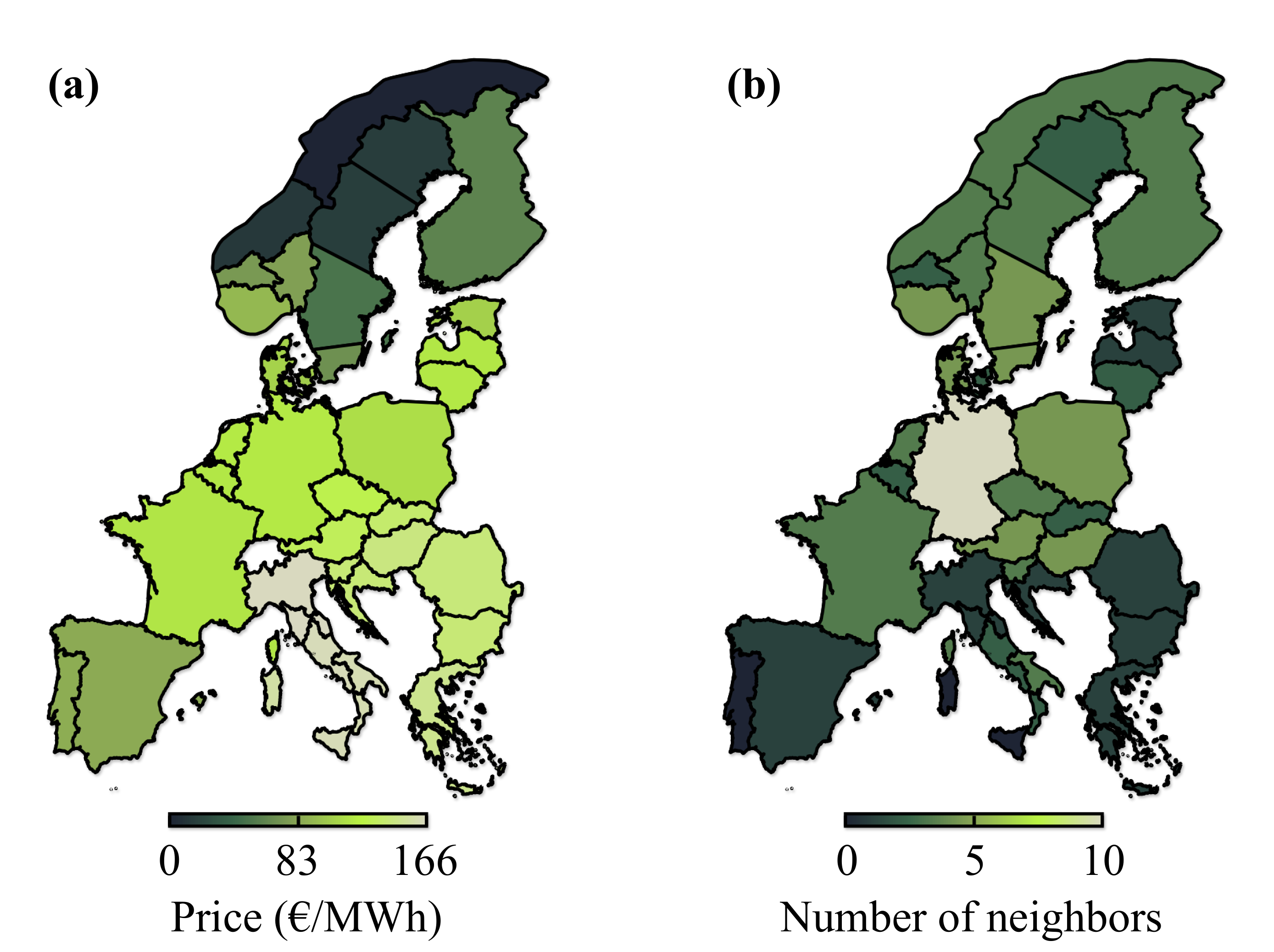}
\caption{
Spatial distribution of electricity price and number of neighboring regions. 
    \textbf{(a)} Electricity prices for 38 European regions averaged from 2022-01-01 to 2026-01-01. A significant zonal price difference is observed between north and south regions.
    \textbf{(b)} Number of neighboring regions that are \emph{directly} connected to certain region via transmission lines. For example, France and Portugal  are connected to Spain, thus the number of neighboring regions for Spain is 2. The mean value across all regions is 3.4.}
    \label{fig:spatial_visualization}
\end{figure} 
can restrict electricity flow between regions and lead to zonal price differences~\cite{finck2021impact}, 
illustrated in Figure~\ref{fig:spatial_visualization}. 
These price disparities highlight the spatial nature of electricity price formation. 
Recent studies show that electricity price dynamics are strongly influenced by spatial interdependencies and cannot be accurately captured using region-specific models \cite{do2024electricity, yu2026deeplearningelectricityprice}. Therefore, explicitly modeling the spatial structure of the European electricity market is essential for producing accurate price forecasts.

Most existing studies on electricity price forecasting do not explicitly model the spatial structure and focus on a single-region market, particularly Germany 
\cite{MUNIAIN20201193,maciejowskaEnhancingLoadWind2021,kitsatoglouEnsembleApproachEnhanced2024}, 
as the German market is one of the largest markets in Europe. Other studies explore forecasting methods for markets such as Denmark, Finland, Spain, and Austria, also using region-specific models 
\cite{zielDayaheadElectricityPrice2018,loizidisElectricityMarketPrice2024, yu2025orderbookfeaturelearningasymmetric}. 
More recent works model the spatial nature of the electricity price. For instance, a Graph Convolutional Network (GraphConv) is applied to capture spatial interdependencies in the Nordic markets, such as Norway, Sweden, and Finland \cite{yang2024forecasting}. An attention-based variant is developed to predict prices in certain European markets such as Austria, Germany, and Hungary \cite{meng2024day}. 
However, these models cover only subsets of Europe and learn spatial dependencies through fully learnable mechanisms (e.g., spatial convolutions or self-attention). Such designs may inadvertently incorporate signals from topologically distant regions that are weakly related to the target region, introducing noise and increasing the overfitting risk. This motivates incorporating transmission-topology graph knowledge as an explicit regularization to constrain spatial information flow.

Unlike conventional forecasting models trained from scratch, time-series foundation models have achieved remarkable success across diverse domains such as weather,
\begin{figure*}[!t]
\hspace*{-1.8mm}
\centering\includegraphics[width=1.01\textwidth]{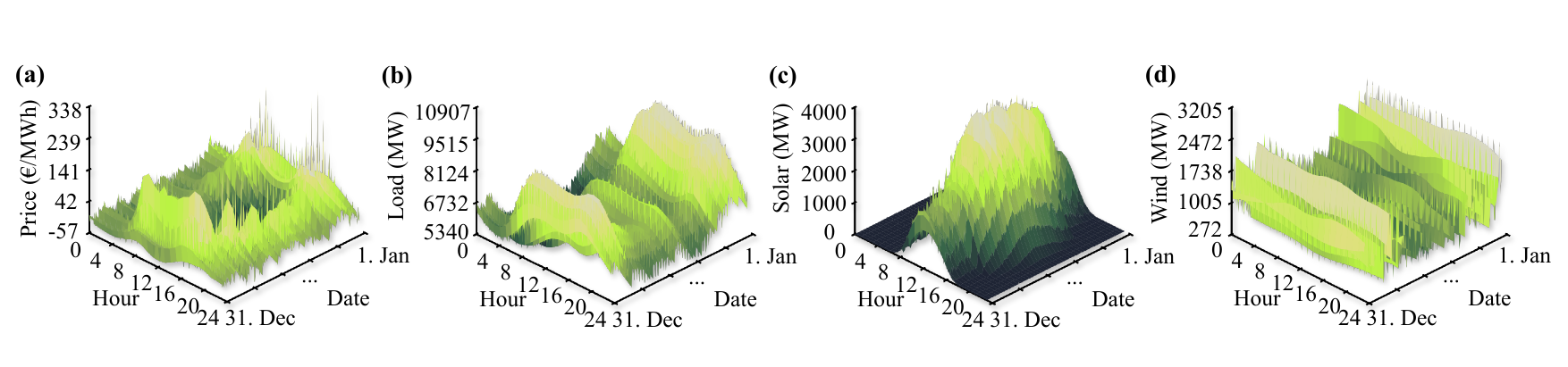}
    \caption{European-level energy data in 2025, averaged across regions.
    \textbf{(a)} Electricity price. Price spikes sharply during the morning and evening peak, dip around midday, and shows higher volatility in the winter.  
    \textbf{(b)} Forecasted load. Load exhibits a double‐peak each day with winter peaks substantially larger than summer.  
    \textbf{(c)} Forecasted solar power generation. Solar is zero overnight, rises in a smooth bell curve to a strong midday maximum, then falls back to zero by dusk, and is much higher in summer.
    \textbf{(d)} Forecasted wind power generation (onshore and offshore). Wind lacks a daily pattern, fluctuates with high‐frequency spikes, and is much higher in winter.}
\label{fig:energy_data_2024}
\end{figure*}
\begin{figure}[!t]
\includegraphics[width=1\linewidth]{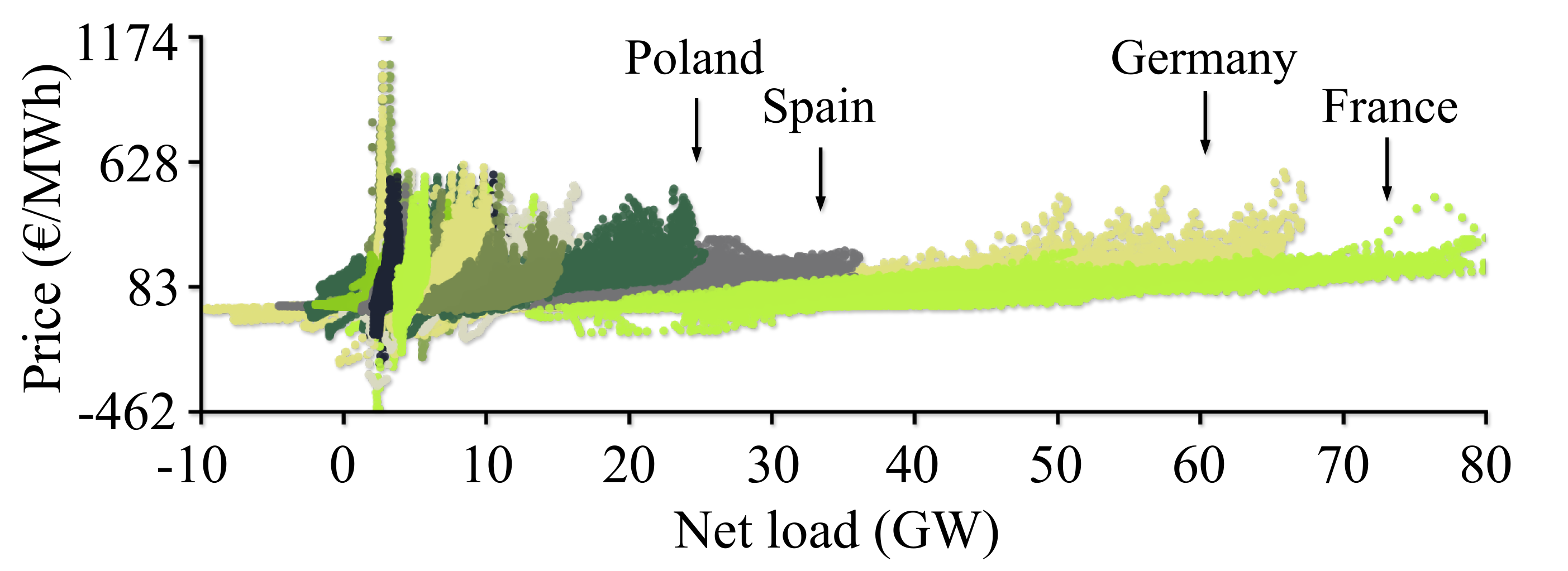}
\caption{
Price–net load relationship across European regions ($\text{Net load} =\text{load} -\text{solar} - \text{wind}$). 
Distinct regional behaviors are evident in markets such as Germany, France, and Spain, while some regions share similar patterns.
}
    \label{fig:price_netload}
\end{figure} 
transportation, and electricity, by capturing complex temporal patterns and exhibiting strong generalization capabilities~\cite{ansari2024chronos, timesfm, liu2025moiraimoe, timemoe}. 
However, electricity prices are shaped not only by local fundamentals but also by signals from neighboring regions through the transmission lines. Existing foundation models combine time series through purely data-driven mixing without incorporating transmission-topology priors, and therefore cannot exploit physically meaningful spatial structure. This gap motivates a domain-specific foundation model with injected graph prior.

To support the development of a domain-specific foundation model for electricity price forecasting, there is a pressing need for high-quality, large-scale, and up-to-date datasets that reflect the spatiotemporal complexity of  integrated European markets. However, existing datasets are often fragmented in structure, cover only short time periods, are outdated, or focus on individual regions \cite{lago2021forecasting}.  This lack of standardized data poses a significant barrier to training and evaluating domain-specific foundation models. We address these limitations by  introducing a comprehensive and up-to-date dataset and proposing PriceFM, a foundation model that utilizes a shared MoE projection layer to process multi-region inputs and regularizes noisy signals from distant regions via a topology-guided sparse graph mask. In summary, our contributions are as follows:
\begin{itemize}
    \item We process and introduce a comprehensive and up-to-date \textcolor{black}{renewable-aware} dataset. To the best of our knowledge, this is the largest and most diverse open dataset for European electricity markets, comprising day-ahead electricity prices, day-ahead forecasts of load, solar, and wind power generation, covering 24 European countries (38 regions), spanning from 2022-01-01 to 2026-01-01.
    \item We propose and release PriceFM, a novel probabilistic forecasting framework that integrates prior graph knowledge derived from the transmission topology of the European electricity market. PriceFM supports multi-region, multi-timestep, and multi-quantile forecasting \textcolor{black}{for renewable-aware electricity market operation}.
    \item We conduct experiments to evaluate the model’s performance against multiple baselines, and assess the impact of design choices through ablation studies, thereby providing both quantitative evidence of overall performance and insights into optimal configurations.
\end{itemize}

\section{Related Work}

\textbf{Generic Foundation Models.}
Foundation models are typically pretrained on large-scale datasets. Representative examples include {Chronos}~\cite{ansari2024chronos}, {TimesFM} ~\cite{timesfm},  {Moirai}~\cite{liu2025moiraimoe}, and TimeMoE~\cite{timemoe}. Pretraining enables these models to learn reusable temporal representations and to generalize across domains without retraining on the target dataset.

\textbf{Time-Series Models.}
Time-series models can be trained from scratch and applied across a wide range of time-series tasks. Representative examples include {FEDFormer}~\cite{zhou2022fedformer}, {iTransformer}~\cite{itransformer}, {PatchTST}~\cite{PatchTST}, {TimesNet}~\cite{TimesNet}, and {TimeXer}~\cite{TimeXer}. Although these methods are not necessarily pretrained as foundation models, they often achieve strong performance when trained end-to-end on the target dataset.

\textbf{Graph Models.}
Graph-based models represent spatial structure by modeling regions as nodes and their relations as edges, enabling information propagation across the graph. Representative examples include {GraphConv~\cite{gcn}, {Graph Attention Network (GraphAttn)}~\cite{gat}, {GraphSAGE}~\cite{GraphSAGE}, {GraphDiffusion}~\cite{GraphDiff}, and {GraphARMA}~\cite{graph_arma}. By incorporating an adjacency matrix, these models learn spatial mixing patterns with temporal dynamics. This property makes them suitable baselines for evaluating whether injecting a topology-constrained sparse graph prior improves multi-region forecasting.

\section{Preliminary}
\label{sec:preliminary}
As we focus on the day-ahead market, the forecasting target is a {probabilistic price trajectory}, i.e., $\mathcal{T}=96$ quarter-hourly prices for the delivery day $\mathcal{D}+1$ with a set of quantiles (\( \tau \in \mathcal{Q} = \{0.10, 0.25, 0.45, 0.50, 0.55, 0.75, 0.90\} \)), using data available before gate closure, typically around midday on day \(\mathcal{D}\). After midday on \(\mathcal{D}\), the electricity prices for \(\mathcal{D}+1\) are published and known.
We employ a backward-looking window of size \({L}\) (e.g. \({L} = 96 \) corresponds to 24 hours from \(\mathcal{D}\)), for known electricity prices, denoted as \(
\mathbf{X}^{\mathrm{price}}_{r_{\mathrm{in}}} \). 

We also include forward-looking exogenous features, such as day-ahead forecasts of {load}, {solar}, and {wind} (sum of onshore and offshore) power generation for \(\mathcal{D}+1\), denoted as \(
\mathbf{X}^{\mathrm{exo}}_{r_{\mathrm{in}}} \), made on \(\mathcal{D}\) before gate closure, as well as their historical values over \({L}\). 
\textcolor{black}{These exogenous variables are directly related to renewable-driven net-load dynamics and are therefore critical for modeling price formation under high renewable penetration.}
The forecasting setup and the choice of feature set are widely used in prior works \cite{maciejowskaAssessingImpactRenewable2020,uniejewskiRegularizedQuantileRegression2021,meng2024day}. 
Importantly, this work utilizes multi-region inputs to produce multi-region, multi-timestep, and multi-quantile forecasts. 
Therefore, the input and output of {PriceFM} are defined as:
\begin{itemize}
  \item \textbf{Input:}  $\left\{\mathbf{X}^{\mathrm{price}}_{r_{\mathrm{in}}},\mathbf{X}^{\mathrm{exo}}_{r_{\mathrm{in}}}\right\}_{r_{\mathrm{in}}\in\mathcal{R}}$, \\ where
  \(
\mathbf{X}^{\mathrm{price}}_{r_{\mathrm{in}}}
    \in \mathbb{R}^{L \times 1}
  \)
  and
  \(
\mathbf{X}^{\mathrm{exo}}_{r_{\mathrm{in}}}
    \in \mathbb{R}^{(L + \mathcal{T})\times 3},
  \)
  \item \textbf{Output:} $\left\{\hat{\mathbf{y}}_{r_{\mathrm{out}},\tau}\right\}_{r_{\mathrm{out}}\in\mathcal{R},\,\tau\in\mathcal{Q}}$, \\
  where $\hat{\mathbf{y}}_{r_{\mathrm{out}},\tau}\in\mathbb{R}^{\mathcal{T}}$,
\end{itemize}
where $r_{\mathrm{in}}, r_{\mathrm{out}}\in\mathcal{R}=\{\text{AT},\dots,\text{SK}\}$ are region codes.

\section{Data}
\label{sec:data}

\subsection{Spatiotemporal Coverage}
\label{sec:coverage}

Spatially, the dataset covers {24 European countries (38 regions)}.
These regions reflect transmission zones rather than administrative boundaries. For example, Denmark (DK) is split into two regions: DK1 and DK2. Each is connected to different regions, resulting in distinct cross-border power flows. 
Temporally, the dataset spans {from 2022-01-01 to 2026-01-01}, providing wide temporal coverage. In total, the dataset contains approximately \textbf{5.3 million} records, making it suitable for foundation model training.

\subsection{Feature Set}
\label{sec:featurechoice}

The feature set includes day-ahead electricity prices, load forecasts, and solar and wind power generation forecasts, where the wind feature is computed by summing the offshore and onshore wind power generation. 
For simplicity, we refer to these features as \textit{price}, \textit{load}, \textit{solar}, and \textit{wind}, respectively.  
A European-level visualization of these features is shown in Figure~\ref{fig:energy_data_2024}.
\textcolor{black}{By explicitly including solar and wind generation forecasts, the dataset captures renewable-driven variability that is central to sustainable power-system operation~\cite{7973035}.}

\subsection{Resolution}
\label{sec:resolution}
We resample all features to a 15-min resolution for two reasons: (1) an increasing number of EU electricity markets are moving from 60-min resolution to 15-min resolution; and (2) the raw data exhibit heterogeneous temporal resolutions. For example, load in Spain is provided hourly before 2022-05-23 and then switches to quarter-hourly resolution afterward;
in Austria, load is reported quarter-hourly while prices are with hourly resolution before 2025-10-01.

\subsection{Missing Value}
\label{sec:missingdata}

Partial features are excluded due to the high rate of  missing values ($>20\%$).
For example, 
solar from Latvia has a 56.6\% missing rate and is only available after 2024-04-07.
The features with low missing rates ($<1\%$) are filled using linear interpolation.
If a region does not provide a certain generation type (e.g., wind), we keep the input dimensionality fixed by adding an all-zero feature, indicating no generation.

\section{PriceFM}
\label{sec:model}

\begin{figure*}[!ht]\centering
\hspace*{-13mm}
\includegraphics[width=1.05\linewidth]{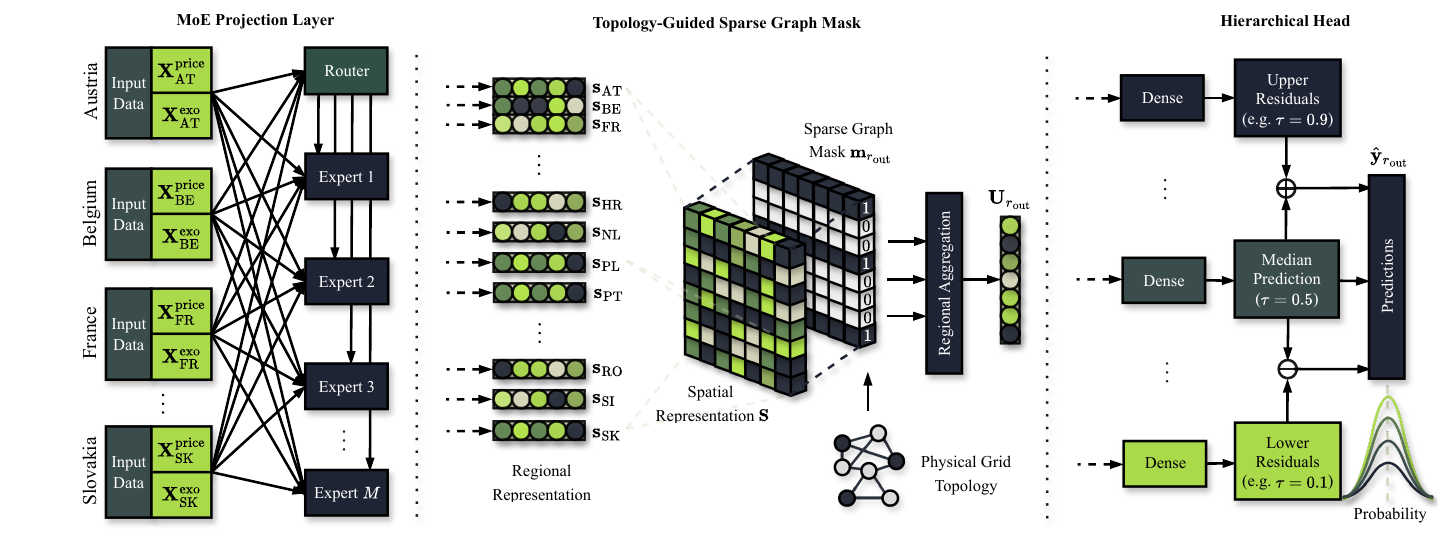}
\caption{Structure of PriceFM. The input features \( \mathbf{X}^{\mathrm{price}}_{r_{\mathrm{in}}} \) and \( \mathbf{X}^{\mathrm{exo}}_{r_\mathrm{in}} \) are passed into a MoE projection layer to produce the regional representations. 
The regional representations are  stacked to form the shared spatial representation \( \mathbf{S} \), which is multiplied with the sparse graph mask to produce the spatial representation \( \mathbf{U}_{r_{\mathrm{out}}} \).
\( \mathbf{U}_{r_{\mathrm{out}}} \) is fed into hierarchical quantile heads to produce probabilistic forecasts.} \label{fig:model} \end{figure*}

\subsection{MoE Projection Layer}
\label{sec:moeprojection}

As later introduced in Section~\ref{sec:adjmatrixformation}, we will inject graph knowledge to compute price representations across regions. This requires that the regional price representations are \emph{comparable} and lie in a shared embedding space. 
A natural solution is to assign  38 dense layers to 38 input regions. However, as shown in Fig.~\ref{fig:price_netload}, some regions exhibit {similar patterns}, suggesting that they can {share} parts of the projection mechanism. To this end, we design a \emph{shared} Mixture-of-Experts (MoE) projection layer that maps each region's inputs $(\mathbf{X}^{\mathrm{price}}_{r_{\mathrm{in}}}, \mathbf{X}^{\mathrm{exo}}_{r_{\mathrm{in}}})$ into a regional representation.

\textbf{Fusion Expert.}
We reshape each modality into a latent embedding of dimension $h$ via a dense layer with \emph{Swish} activation, and inject the exogenous representation as a residual into the price representation:
\begin{align}
\mathbf{X}^{\mathrm{price}}_{r_{\mathrm{in}}}
&\xrightarrow{\text{Project}}
\hat{\mathbf{z}}^{\mathrm{price}}_{r_{\mathrm{in}}}
\in \mathbb{R}^{h}, \\
\mathbf{X}^{\mathrm{exo}}_{r_{\mathrm{in}}}
&\xrightarrow{\text{Project}}
\hat{\mathbf{z}}^{\mathrm{exo}}_{r_{\mathrm{in}}}
\in \mathbb{R}^{h}, \\
\mathbf{z}_{r_{\mathrm{in}}}
&=\mathrm{Swish}\!\left(
\hat{\mathbf{z}}^{\mathrm{price}}_{r_{\mathrm{in}}}
+
\hat{\mathbf{z}}^{\mathrm{exo}}_{r_{\mathrm{in}}}\right)
\in \mathbb{R}^{h}.
\label{eq:moe_residual_fusion_vec}
\end{align}

\textbf{Weighting Router.}
Similar to the fusion expert in Eq.~\eqref{eq:moe_residual_fusion_vec}, the router takes the same pair of inputs $(\mathbf{X}^{\mathrm{price}}_{r_{\mathrm{in}}}, \mathbf{X}^{\mathrm{exo}}_{r_{\mathrm{in}}})$, but uses a dense layer with \emph{softmax} activation to output the expert weights:
\begin{align}
\mathbf{X}^{\mathrm{price}}_{r_{\mathrm{in}}}
&\xrightarrow{\text{Project}}
\hat{\boldsymbol{\pi}}^{\mathrm{price}}_{r_{\mathrm{in}}}
\in \mathbb{R}^{M}, \\
\mathbf{X}^{\mathrm{exo}}_{r_{\mathrm{in}}}
&\xrightarrow{\text{Project}}
\hat{\boldsymbol{\pi}}^{\mathrm{exo}}_{r_{\mathrm{in}}}
\in \mathbb{R}^{M}, \\
\boldsymbol{\pi}_{r_{\mathrm{in}}}
&=
\mathrm{Softmax}\!\left(
\hat{\boldsymbol{\pi}}^{\mathrm{price}}_{r_{\mathrm{in}}}
+
\hat{\boldsymbol{\pi}}^{\mathrm{exo}}_{r_{\mathrm{in}}}
\right)
\in \mathbb{R}^{M}.
\label{eq:moe_gating}
\end{align}

Let $M$ denote the number of experts. Given $\mathbf{z}_{r_{\mathrm{in}}}\in\mathbb{R}^{h}$, the $M$ experts output an \emph{expert matrix}:
\begin{equation}
\mathbf{Z}_{r_{\mathrm{in}}}
=
\begin{bmatrix}
(\mathbf{z}_{r_{\mathrm{in}}})_{1} \\
(\mathbf{z}_{r_{\mathrm{in}}})_{2} \\
\vdots \\
(\mathbf{z}_{r_{\mathrm{in}}})_{M}
\end{bmatrix}
\in \mathbb{R}^{M\times h},
\label{eq:moe_expert_matrix}
\end{equation}
where each row $(\mathbf{z}_{r_{\mathrm{in}}})_{m}\in\mathbb{R}^{h}$ is the output embedding produced by fusion expert $m$.

The output of the MoE projection layer is then computed in vectorized form as:
\begin{equation}
\mathbf{S}_{r_{\mathrm{in}}}
=
\boldsymbol{\pi}_{r_{\mathrm{in}}}^{\top}\mathbf{Z}_{r_{\mathrm{in}}}
\in \mathbb{R}^{h}.
\label{eq:moe_projection_output}
\end{equation}

\subsection{Topology-Guided Sparse Graph Mask}
\label{sec:adjmatrixformation}

As electricity markets are physically coupled through cross-border transmission lines, this motivates a topology-aware modeling prior: input regions that are topologically { closer} to the target region $r_{\mathrm{out}}$ typically exert a {stronger} impact than regions that are farther away.
Incorporating features from distant regions can introduce irrelevant or noisy signals, harming the generalization of the model. 
To explicitly encode this structure, we design a topology-guided graph mask to construct a sparse, output-region-specific connectivity pattern for aggregating regional representations.

\textbf{Graph Distance.}
We produce \emph{graph distance} 
by performing a breadth-first search (BFS) traversal on the cross-border grid topology, 
detailed in Appendix, Table \ref{tab:neighbors}.
For a given output region \( r_{\mathrm{out}} \in \mathcal{R} \), we define the graph distance \( d(r_{\mathrm{in}}, r_{\mathrm{out}}) \) as the minimal number of transmission hops from each input region \( r_{\mathrm{in}} \) to the output region \( r_{\mathrm{out}} \), based on direct or indirect physical connectivity: 
\begin{equation}
d(r_{\mathrm{in}}, r_{\mathrm{out}}) =
\begin{cases}
0 & \text{if } r_{\mathrm{in}} = r_{\mathrm{out}}, \\
1 & \text{if } r_{\mathrm{in}} \sim r_{\mathrm{out}}, \\
1 + \min\limits_{r' \sim r_{\mathrm{in}}} d(r', r_{\mathrm{out}}) & \text{otherwise},
\end{cases}
\label{eq:bfs_distance}
\end{equation}
where \( r_{\mathrm{in}} \sim r_{\mathrm{out}} \) denotes that two regions are directly connected by a transmission line. For example, let \( r_{\mathrm{out}} = \text{AT} \). Then \( d(\text{AT}, \text{AT}) = 0 \). The region HU is directly connected to AT, thus \( d(\text{HU}, \text{AT}) = 1 \). SK is indirectly connected to AT via HU, yielding \( d(\text{SK}, \text{AT}) = 2 \).

\textbf{Sparse Graph.}

\textcolor{black}{If a region experiences a renewable-driven event, such as a surge in solar or wind generation,} its impact will first affect its neighborhood and then propagate gradually along the topology before reaching the neighborhood of the target region. 
Being said, to model the target region accurately, we should {prioritize} input features from its {closer} neighbors. 

Motivated by this propagation mechanism and the observation that distant features may be noisy, we construct a sparse graph mask to restrict information flow to a bounded neighborhood of each target region.
The mask provides a \textbf{binary control} mechanism: if a region is assigned a value of {1}, its feature representation is {retained} for aggregation; otherwise, it is excluded.
The \textbf{relative importance} among the retained regions is then learned in a \textbf{data-driven} manner.

Specifically, for each target region $r_{\mathrm{out}}\in\mathcal{R}$, we compute the graph distance $d(r_{\mathrm{in}}, r_{\mathrm{out}})$ for all input regions $r_{\mathrm{in}}\in\mathcal{R}$ using Eq.~\eqref{eq:bfs_distance} and define the output-specific mask:
\begin{equation}
\mathbf{m}_{r_{\mathrm{out}}}
=
\begin{bmatrix}
\mathbb{I}\!\left(d(\mathrm{AT}, r_{\mathrm{out}})\le \delta\right) \\
\mathbb{I}\!\left(d(\mathrm{BE}, r_{\mathrm{out}})\le \delta\right) \\
\vdots \\
\mathbb{I}\!\left(d(\mathrm{SK}, r_{\mathrm{out}})\le \delta\right)
\end{bmatrix}
\in \{0,1\}^{|\mathcal{R}|\times 1},
\label{eq:sparse_mask_vector}
\end{equation}
where $\mathbb{I}(\cdot)$ is the indicator function, and $\delta\in\mathbb{N}$ is the graph degree cutoff controlling the maximum neighborhood radius retained for $r_{\mathrm{out}}$. By controlling $\delta$, we can perform case studies for each target region to understand {how far} along the grid topology neighboring information {remains beneficial}.

As an example, let $r_{\mathrm{out}}=\mathrm{AT}$ and $\delta=0$. Then, only AT is assigned a mask value of $1$, the rest input regions are assigned $0$, meaning that no information from any neighbors is used. If $\delta=1$, then only regions directly connected to AT are assigned a mask value of $1$ (e.g., HU and SI), while all other regions with $d(r_{\mathrm{in}},\mathrm{AT})>1$ are assigned $0$.

\textbf{Regional Aggregation.}
The regional embeddings $\{\mathbf{S}_{r_{\mathrm{in}}}\}_{r_{\mathrm{in}}\in\mathcal{R}}$ from Eq. (\ref{eq:moe_projection_output}) are stacked to form the spatial representation:
\begin{equation}
\mathbf{S}
=
\mathrm{Stack}\!\left(\left\{\mathbf{S}_{r_{\mathrm{in}}}\right\}_{r_{\mathrm{in}}\in\mathcal{R}}\right)
\in \mathbb{R}^{|\mathcal{R}|\times h}.
\label{eq:spatialrepres_moe}
\end{equation}

The topology-guided sparsity is injected into $\mathbf{S}$ by computing a sparsity-constrained average representation over the masked neighborhood of each target region $r_{\mathrm{out}}$:
\begin{equation}
\mathbf{U}_{r_{\mathrm{out}}}
=
\frac{
\mathbf{m}_{r_{\mathrm{out}}}^\top \mathbf{S}
}{
\mathbf{m}_{r_{\mathrm{out}}}^\top \mathbf{1}
},
\label{eq:sparse_avg_vectorized}
\end{equation}
where $\mathbf{1}\in\mathbb{R}^{|\mathcal{R}|\times 1}$ is a vector of ones. After multiplication with the mask, regional inputs corresponding to a mask value of {0} make {no contribution} to the aggregated representation.
This operation acts as \textbf{spatial regularization} by restricting aggregation to a physically plausible neighborhood.

\subsection{Hierarchical Head}
\label{sec:head}
To prevent quantile crossing\footnote{Quantile crossing refers to the phenomenon where upper quantile predictions (e.g., 90\%) fall below lower quantiles (e.g., 10\%), violating the monotonicity of the quantile function.}, we modify the hierarchical quantile head proposed in \cite{yu2025orderfusion} for multi-region, multi-timestep, and multi-quantile forecasting tasks.

Specifically, the median quantile (\( \tau_m = 0.5 \)) price trajectory, which represents the full set of timesteps \( \mathcal{T} \), is predicted from \( \mathbf{U}_{r_{\mathrm{out}}} \) via a dense layer \( \mathcal{F}_{\tau_m}(\cdot) \):
\begin{equation}
    \hat{\mathbf{y}}_{{r_{\mathrm{out}}}, \tau_m } = \mathcal{F}_{\tau_m}(\mathbf{U}_{r_{\mathrm{out}}}) \in \mathbb{R}^{\mathcal{T}}.
\end{equation}

To produce the upper quantile forecast (\( \tau_u > 0.50 \)), a residual price trajectory \( \hat{\mathbf{r}}_{{r_{\mathrm{out}}}, \tau_u} \) is generated from \( \mathbf{U}_{r_{\mathrm{out}}} \):
\begin{equation}
    \hat{\mathbf{r}}_{{r_{\mathrm{out}}}, \tau_u} = \mathcal{F}_{\tau_u}(\mathbf{U}_{r_{\mathrm{out}}}) \in \mathbb{R}^{\mathcal{T}},
    \label{eq:res_upper}
\end{equation}
where a non-negative function \( g(\cdot) \), such as absolute-value function, is applied to the price residual. The final upper quantile forecast is obtained by adding this non-negative residual to the median:
\begin{equation}
    \hat{\mathbf{y}}_{{r_{\mathrm{out}}}, \tau_u} = \hat{\mathbf{y}}_{{r_{\mathrm{out}}}, \tau_m} + g(\hat{\mathbf{r}}_{{r_{\mathrm{out}}}, \tau_u}).
    \label{eq:upper_add}
\end{equation}
For the lower quantile (\( \tau_l < 0.50 \)), we compute a residual trajectory similarly:
\begin{equation}
    \hat{\mathbf{r}}_{{r_{\mathrm{out}}}, \tau_l} =  \mathcal{F}_{\tau_l}(\mathbf{U}_{r_{\mathrm{out}}}) \in \mathbb{R}^{\mathcal{T}},
    \label{eq:res_lower}
\end{equation}
and subtract it from the median to obtain the lower quantile prediction:
\begin{equation}
    \hat{\mathbf{y}}_{{r_{\mathrm{out}}}, \tau_l} = \hat{\mathbf{y}}_{{r_{\mathrm{out}}}, \tau_m} - g(\hat{\mathbf{r}}_{{r_{\mathrm{out}}}, \tau_l}).
    \label{eq:lower_add}
\end{equation}
This hierarchical design guarantees that the upper quantile prediction is greater than or equal to the lower one at each time step, overcoming quantile crossing.

\subsection{Loss}
\label{sec:loss}

We use the \textit{Average Quantile Loss (AQL)} as the training objective for multi-region, multi-timestep, and multi-quantile probabilistic forecasting. Let \( y_{i, {r_{\mathrm{out}}}, t} \) denote the ground-truth price for the \( i \)-th training sample, output region \( {r_{\mathrm{out}}} \), and timestep \( t \), and let \( \hat{y}_{i, {r_{\mathrm{out}}}, t , \tau} \) be the corresponding predicted quantile. The AQL is computed as:
{\small
\begin{equation}
\text{AQL} = \frac{1}{N \left| \mathcal{R} \right| \mathcal{T} \left| \mathcal{Q} \right|} 
\sum_{i=1}^N \sum_{r_{\mathrm{out}} \in \mathcal{R}} \sum_{t=1}^{\mathcal{T}} \sum_{\tau \in \mathcal{Q}}  
L_\tau\left(y_{i, r_{\mathrm{out}}, t}, \hat{y}_{i, r_{\mathrm{out}}, t, \tau} \right),
\end{equation}
}

where \( N \) is the number of samples, and the quantile loss \( L_\tau \) is defined as:
\begin{equation}
    L_\tau(y, \hat{y}_\tau) = 
    \begin{cases} 
      \tau \cdot (y - \hat{y}_\tau), & \text{if } y \geq \hat{y}_\tau, \\
      (1 - \tau) \cdot (\hat{y}_\tau - y), & \text{otherwise},
    \end{cases}
\end{equation}
where \( y \) and \( \hat{y} \) are the true and predicted values, respectively.

\begin{table}[t]
\begin{center}
\caption{Model Capability Comparison.}
\label{tab:baseline_capabilities}
\begin{tabular}{lcc}
\toprule
\textbf{Model} & \textbf{Multivariate Input} & \textbf{Probabilistic Output} \\
\midrule
Chronos              &  & $\checkmark$ \\
Chronos$^{[2.0]}$        & $\checkmark$ & $\checkmark$ \\
Moirai$^{[\text{S}]}$            & $\checkmark$ & $\checkmark$ \\
Moirai$^{[\text{M}]}$           & $\checkmark$ & $\checkmark$ \\
Moirai$^{[\text{L}]}$            & $\checkmark$ & $\checkmark$ \\
TimeMoE              &  &  \\
TimesFM$^{[2.0]}$         &  & $\checkmark$ \\
TimesFM$^{[2.5]}$        & $\checkmark$ & $\checkmark$ \\
PriceFM              & $\checkmark$ & $\checkmark$ \\
\bottomrule
\end{tabular}
\end{center}
\end{table}

\section{Baselines}

We compare PriceFM with four categories of models, including \textbf{Na\"ive Models}, \textbf{Generic Foundation Models}, \textbf{Time-Series Models}, and \textbf{Graph Models}.

\subsection{Na\"ive Models}

We include na\"ive baselines as reference models, where only historical prices are used as input:
\textbf{Na\"ive$^1$} uses 96 prices from the previous day;
\textbf{Na\"ive$^2$} uses 96 prices averaged over the past three days;
and \textbf{Na\"ive$^3$} uses 96 prices averaged over the past seven days.
To obtain probabilistic forecasts, we compute empirical quantiles at individual levels for each delivery hour.
Seasonal na\"ive baselines are commonly used to evaluate the autoregressive strength of the signal and often serve as strong baselines \cite{zielDayaheadElectricityPrice2018, lago2021forecasting}.

\subsection{Time-Series Models}

To investigate whether graph topology is beneficial, we include several advanced pure time-series models:
\textbf{FEDFormer}, \textbf{iTransformer}, \textbf{PatchTST}, \textbf{TimesNet}, and \textbf{TimeXer}.
Since these models do not support graph-structured inputs, regional input features are concatenated along the feature dimension.
As not all regional inputs are necessarily useful and some may introduce noisy signals, this model category is used to examine the performance degradation caused by unconstrained regional feature mixing.

\subsection{Graph Models}

We also include several graph models to examine whether graph knowledge alone is sufficient without the proposed sparse design.
The graph baselines include
\textbf{GraphConv}, \textbf{GraphAttn}, \textbf{GraphSAGE}, \textbf{GraphDiffusion}, and \textbf{GraphARMA}.
The adjacency matrix is described in Appendix~\ref{spatial_adj}.
Although these models encode regional connectivity through the adjacency matrix, they do not explicitly enforce a binary, target-region-specific information-flow constraint.
Therefore, noisy signals from distant regions may still be assigned weights.

\subsection{Generic Foundation Models}

There exist multiple generic foundation models, including
\textbf{Chronos} (original and 2.0), \textbf{Moirai} (small, base, and large), \textbf{TimesFM} (2.0 and 2.5), and \textbf{TimeMoE}. 
The model capabilities are summarized in Table~\ref{tab:baseline_capabilities}. For models that support multivariate inputs, we use the same input features as PriceFM; otherwise, only historical prices are used as input. 
\textcolor{black}{These models are pretrained on diverse datasets, including weather, transportation, and also European \textbf{electricity} datasets \cite{ansari2024chronos, timesfm}. }
It is therefore important to investigate whether generic pretraining is sufficient to match the proposed method, especially when:
(1) the training data contain not only electricity data but also data from other domains, raising the question of whether such broad-domain pretraining can improve electricity price forecasting accuracy; 
(2) the models are not equipped with graph priors, which further raises the question of whether these models can capture the spatial dependencies implicitly.

To investigate this, we apply a \textbf{leave-one-region-out zero-shot} strategy, i.e., PriceFM is pretrained on 37 regions and directly evaluated on the remaining unseen region.
\textcolor{black}{This leave-one-region-out generalization setting is important, as different regions can exhibit substantially different dynamics, for example, dynamics primarily caused by \textbf{wind} power generation in the Netherlands versus dynamics primarily caused by \textbf{solar} power generation in Spain.}
This comparison can guide the selection of pretrained models for future zero-shot applications in unseen regions, where full-shot training is limited.

\section{Experiments}
\label{sec:exp}

\subsection{Experimental Settings}

\textbf{Rolling Evaluation.}
We adopt a 3-fold rolling evaluation. In fold~1, the data span from 1.~Jan~2022 to 1.~Sep~2024 for training, 1.~Sep~2024 to 1.~Jan~2025 for validation, and 1.~Jan~2025 to 1.~May~2025 for testing. Each subsequent fold advances by 4 months, ending at 1.~Jan~2026, so that the testing windows jointly cover one full year.

\textbf{Data Scaling.} To scale the data while being robust to extreme values, we employ a \texttt{RobustScaler} fitted on the training data, using \texttt{Scikit-Learn}. The fitted scaler is then used to transform validation and testing data. 

\textbf{Evaluation Metrics.}
To evaluate the probabilistic performance, we utilize AQL and Average Quantile Crossing Rate (AQCR). 
For pointwise forecasting, we use Mean Absolute Error (MAE) and Root Mean Squared Error (RMSE).

\subsection{Empirical Results}

\textbf{Against Na\"ive Models.}
As shown in Table~\ref{tab:full_shot_compact}, PriceFM substantially outperforms the na\"ive baselines across both probabilistic and pointwise metrics.
The units of AQL, RMSE, and MAE are expressed in~\euro{}/\text{MWh}, while AQCR is expressed in~\%.
Compared with the best na\"ive baseline, PriceFM reduces AQL by \textbf{62.1\%}, MAE by \textbf{35.3\%}, and RMSE by \textbf{31.3\%}.
Although the na\"ive baselines naturally achieve zero AQCR due to empirical quantile construction, their much larger AQL and pointwise errors indicate limited forecasting accuracy.

\begin{table}[t]
\centering
\caption{Model Comparison with Full-Shot Evaluation Strategy.}
\label{tab:full_shot_compact}
\begin{tabular}{
>{\raggedright\arraybackslash}p{1.55cm}
>{\centering\arraybackslash}p{0.9cm}
>{\centering\arraybackslash}p{1.1cm}
>{\centering\arraybackslash}p{0.9cm}
>{\centering\arraybackslash}p{1.1cm}
>{\centering\arraybackslash}p{0.9cm}
}
\toprule
\multirow{2}{*}{\textbf{Model}}
& \multicolumn{2}{c}{\cellcolor{valgreen!60}\textbf{Probabilistic}}
& \multicolumn{2}{c}{\cellcolor{valgreen!60}\textbf{Pointwise}}
& \multirow{2}{*}{\textbf{Rank}} \\
\cmidrule(lr){2-3}
\cmidrule(lr){4-5}
& \textbf{AQL} 
& \textbf{AQCR}
& \textbf{MAE} 
& \textbf{RMSE} 
& \\

\midrule

Na\"ive$^{1}$      & 15.29 & \textcolor{gray}{0.00} & 22.06 & 34.68 & 11 \\
Na\"ive$^{2}$      & 15.35 & \textcolor{gray}{0.00} & 23.31 & 34.31 & 13 \\
Na\"ive$^{3}$      & 15.46 & \textcolor{gray}{0.00} & 22.64 & 32.61 & 12 \\

\midrule

FEDFormer          & 8.22  & 15.33 & 20.15 & 31.75 & 8  \\
PatchTST           & 8.06  & 18.21 & 20.20 & 31.59 & 8  \\
iTransformer       & 8.24  & 13.96 & 21.03 & 32.11 & 9  \\
TimesNet           & 7.98  & 13.42 & 19.48 & 30.94 & 7  \\
TimeXer            & 8.30  & 14.77 & 21.94 & 31.88 & 10 \\

\midrule

GraphConv          & 6.61  & 6.88  & 16.81 & 25.97 & 4  \\
GraphAttn          & 7.13  & 10.33 & 17.00 & 26.11 & 6  \\
GraphSAGE          & 6.78  & 6.01  & 17.56 & 26.03 & 5  \\

\rowcolor{relgreen!10}
GraphDiffusion     & 6.69  & 5.72  & 16.44 & 25.93 & 2  \\
GraphARMA          & 6.72  & 6.03  & 16.56 & 25.84 & 3  \\

\midrule

\rowcolor{customgray!20}
\textbf{PriceFM}   & \textbf{5.80} & \textbf{0.00} & \textbf{14.28} & \textbf{22.39} & \textbf{1} \\

\bottomrule
\end{tabular}
\end{table}

\begin{table}[t]
\centering
\caption{Model Comparison with Leave-One-Out Evaluation Strategy.}
\label{tab:zero_shot_compact}
\begin{tabular}{
>{\raggedright\arraybackslash}p{1.55cm}
>{\centering\arraybackslash}p{0.9cm}
>{\centering\arraybackslash}p{1.1cm}
>{\centering\arraybackslash}p{0.9cm}
>{\centering\arraybackslash}p{1.1cm}
>{\centering\arraybackslash}p{0.9cm}
}
\toprule
\multirow{2}{*}{\textbf{Model}}
& \multicolumn{2}{c}{\cellcolor{valgreen!60}\textbf{Probabilistic}}
& \multicolumn{2}{c}{\cellcolor{valgreen!60}\textbf{Pointwise}}
& \multirow{2}{*}{\textbf{Rank}} \\
\cmidrule(lr){2-3}
\cmidrule(lr){4-5}
& \textbf{AQL} 
& \textbf{AQCR} 
& \textbf{MAE}
& \textbf{RMSE} 
& \\

\midrule

Chronos
& 11.14 & 0.00 & 25.98 & 42.19 & 3 \\

\rowcolor{relgreen!10}
Chronos$^{[2.0]}$ 
& 8.03 & 0.00 & 19.44 & 30.93 & 2 \\

Moirai$^{[\text{S}]}$ 
& 11.24 & 0.00 & 27.22 & 43.66 & 4 \\

Moirai$^{[\text{M}]}$
& 12.07 & 0.00 & 30.47 & 47.94 & 6 \\

Moirai$^{[\text{L}]}$
& 11.94 & 0.00 & 29.13 & 46.66 & 5 \\

TimeMoE
& -- & -- & 25.54 & 40.83 & -- \\

TimesFM$^{[2.0]\text{*}}$
& 10.50 & 0.00 & 26.01 & 41.91 & 3 \\

\rowcolor{relgreen!10}
TimesFM$^{[2.5]\text{*}}$
& 7.97 & 0.00 & 19.48 & 30.83 & 2 \\

\midrule
\rowcolor{customgray!20}
\textbf{PriceFM}$^{\text{*}}$
& \textbf{6.91}
& \textbf{0.00}
& \textbf{16.90}
& \textbf{26.24}
& \textbf{1} \\

\rowcolor{customgray!20}
\textbf{PriceFM}
& \textbf{6.85}
& \textbf{0.00}
& \textbf{16.83}
& \textbf{26.13}
& \textbf{1} \\

\bottomrule
\end{tabular}
\end{table}

\textbf{Against Time-Series Models.}
As shown in Table~\ref{tab:full_shot_compact}, PriceFM also clearly outperforms pure time-series models.
The best time-series baseline, TimesNet, still produces \textbf{37.6\%} higher AQL, \textbf{36.4\%} higher MAE, and \textbf{38.2\%} higher RMSE than PriceFM.
This is expected because pure time-series models lack explicit spatial inductive bias. Concatenating multi-region inputs along the feature dimension can introduce spurious correlations and increase the risk of overfitting.

\textbf{Against Graph Models.}
Compared with graph baselines, PriceFM improves AQL by \textbf{12.3\%} over GraphConv, the best-performing graph baseline in terms of probabilistic accuracy.
It also reduces MAE by \textbf{13.1\%} compared with GraphDiffusion and RMSE by \textbf{13.4\%} compared with GraphARMA.
We attribute these gains to the proposed sparse graph masking mechanism, which controls spatial information flow and mitigates the tendency of purely data-driven GNNs to overfit by \textbf{propagating noisy signals} from weakly related regions.
Moreover, while all graph baselines exhibit nonzero quantile crossing, PriceFM achieves zero AQCR, overcoming the crossing problem without sacrificing accuracy.

\textbf{Against Generic Foundation Models.}
Table~\ref{tab:zero_shot_compact} reports the leave-one-out comparison against generic foundation models.
The symbols S, M, and L denote the small, base, and large variants of Moirai.
The symbol $^{\text{*}}$ indicates that TimesFM supports only a fixed set of quantiles
(\( \tau \in \mathcal{Q} = \{0.10, 0.20, 0.30, 0.40, 0.50, 0.60, 0.70, 0.80, 0.90\} \)).
Therefore, PriceFM$^{\text{*}}$ is evaluated using the same quantile set against TimesFM$^{\text{*}}$ for a fair comparison.
Among the non-starred models, PriceFM achieves the best overall performance, reducing AQL by \textbf{14.7\%} compared with the strongest generic foundation model, Chronos$^{[2.0]}$, while also reducing MAE by \textbf{13.4\%} and RMSE by \textbf{15.5\%}.
Under the starred comparison, PriceFM$^{\text{*}}$ also outperforms TimesFM$^{[2.5]\text{*}}$, reducing AQL by \textbf{13.3\%}, MAE by \textbf{13.2\%}, and RMSE by \textbf{14.9\%}.
These results suggest that, even though generic foundation models are pretrained on diverse datasets including electricity data, they remain insufficient for capturing the spatial dependencies of interconnected electricity markets without graph priors.

\section{Ablation Study}
\label{sec:AssessStudy}

\subsection{Spatiotemporal Configurations}

\begin{itemize}
    \item \textbf{Graph Degree Cutoff:} Spatially, we evaluate \(\delta \in \{0, 1, 2, 3, \ldots,  10\}\), ranging from strong constraint to weak constraint. In total, 1,254 trials are conducted to determine the optimal cutoff value for each output region.

    \item \textbf{Backward-Looking Window Size:} Temporally, we compare \(L \in \{96, 288, 672\}\), corresponding to 1 day, 3 days, and 1 week. For each window size, all other hyperparameters are re-optimized.

 \end{itemize}

\textbf{Spatially}, Figure~\ref{fig:spatial_curvature} illustrates the testing loss and the distribution of optimal graph cutoff values.
Some regions confirm the {spatial interdependencies}, as reflected by nonzero optimal cutoff values $(\delta \ne 0)$, while others, such as Germany and France, achieve better performance {without} using neighboring information.
This highlights the distinct spatial behaviors across European regions and shows that the proposed method can {unify} these heterogeneous patterns through region-specific graph cutoff selection.
\textbf{Temporally}, the results in Table~\ref{table_ablation} indicate that the optimal backward-looking window size is \textbf{96}, potentially because information from the distant past becomes outdated. This observation may be related to market efficiency and should be continuously monitored in the future.

\begin{figure}[!ht]
\includegraphics[width=0.97\linewidth]{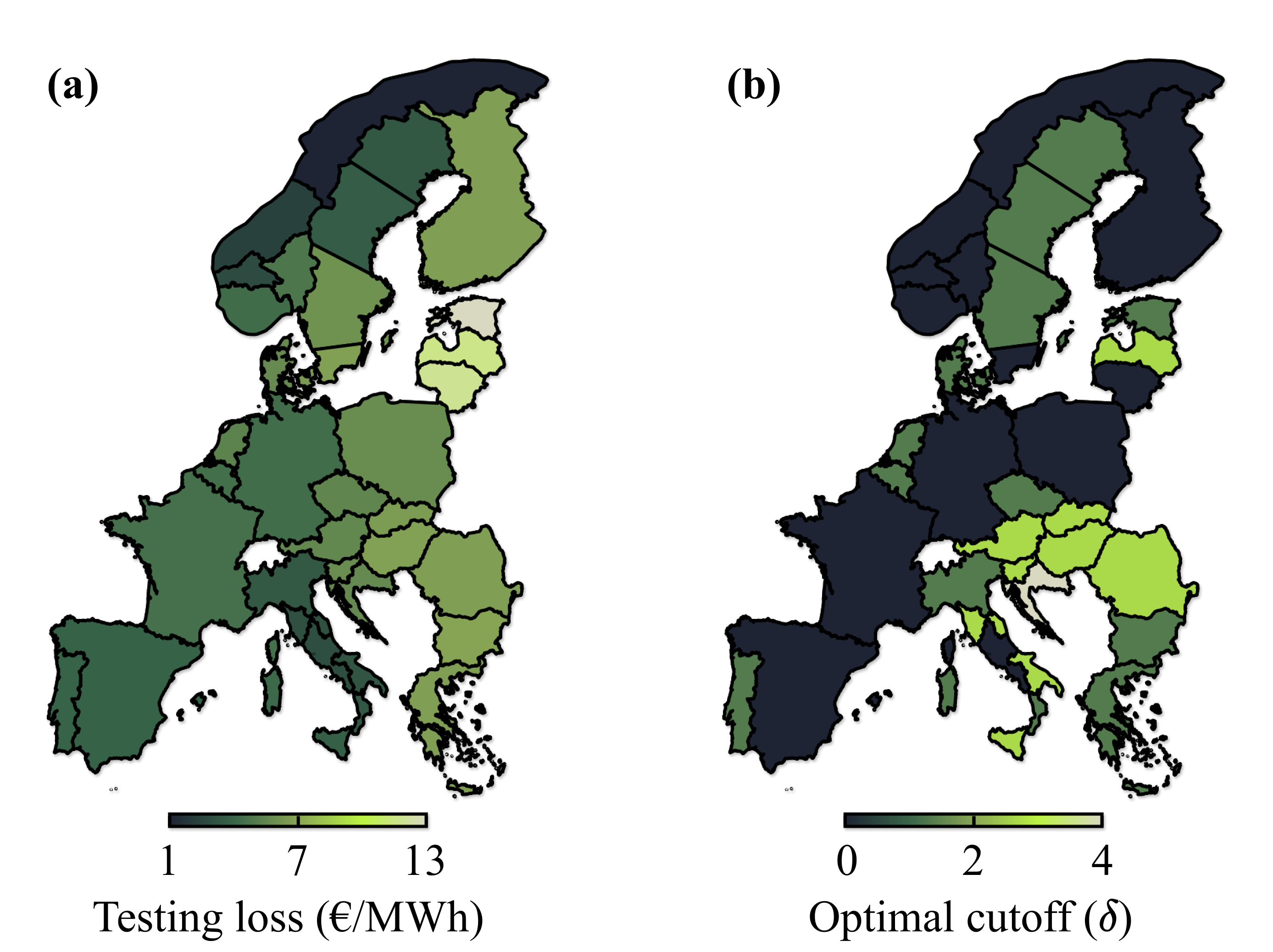}
    \caption{Spatial distribution of testing loss and graph cutoff values. 
    \textbf{(a)} Average quantile loss per region on the testing set. Western and northern European regions exhibit lower losses. 
    \textbf{(b)} Optimal value of graph degree cutoff  per region. 
    Notably, regions such as Germany, France, and Norway have a value of \(0\), indicating optimal performance by excluding neighboring features.}
    \label{fig:spatial_curvature}
\end{figure}

\begin{table}[t]
\centering
\caption{Ablation studies of different module choices.}
\label{table_ablation}
\begin{tabular}{
>{\raggedright\arraybackslash}p{1.55cm}
>{\centering\arraybackslash}p{0.9cm}
>{\centering\arraybackslash}p{1.1cm}
>{\centering\arraybackslash}p{0.9cm}
>{\centering\arraybackslash}p{1.1cm}
>{\centering\arraybackslash}p{0.9cm}
}
\toprule
\multirow{2}{*}{\textbf{Model}}
& \multicolumn{2}{c}{\cellcolor{valgreen!60}\textbf{Probabilistic}}
& \multicolumn{2}{c}{\cellcolor{valgreen!60}\textbf{Pointwise}}
& \multirow{2}{*}{\textbf{Rank}} \\
\cmidrule(lr){2-3}
\cmidrule(lr){4-5}
& \textbf{AQL}
& \textbf{AQCR}
& \textbf{MAE}
& \textbf{RMSE}
& \\
\midrule

\rowcolor{customgray!20}
$L=96$$^{\dagger}$
& \textbf{5.80} & \textbf{0.00} & \textbf{14.28} & \textbf{22.39} & \textbf{1} \\

$L=288$
& 5.86 & {0.00} & 14.30 & 22.51 & 2 \\

$L=672$
& 5.96 & {0.00} & 15.01 & 23.83 & 3 \\

\midrule

$M=1$
& 6.15 & {0.00} & 14.47 & 22.56 & 3 \\

\rowcolor{customgray!20}
$M=4^{\dagger}$
& \textbf{5.80} & \textbf{0.00} & \textbf{14.28} & \textbf{22.39} & \textbf{1} \\

$M=8$
& 5.81 & {0.00} & 14.30 & 22.42 & 2 \\

\midrule

\rowcolor{customgray!20}
Res. Add$^{\dagger}$
& \textbf{5.80} & \textbf{0.00} & \textbf{14.28} & \textbf{22.39} & \textbf{1} \\

Concat.
& 6.11 & {0.00} & 14.80 & 23.03 & 3 \\

Cross-Attn
& {5.79} & {0.00} & 14.33 & 22.41 & 2 \\

\midrule

\rowcolor{customgray!20}
Sparse$^{\dagger}$
& \textbf{5.80} & \textbf{0.00} & \textbf{14.28} & \textbf{22.39} & \textbf{1} \\

Random
& 7.23 & {0.00} & 17.05 & 26.13 & 3 \\

No Mask
& 6.65 & {0.00} & 16.41 & 25.82 & 2 \\

\midrule

\rowcolor{customgray!20}
Absolute$^{\dagger}$
& \textbf{5.80} & \textbf{0.00} & \textbf{14.28} & \textbf{22.39} & \textbf{1} \\

ReLU
& 5.80 & {0.00} & 14.29 & 22.40 & 2 \\

Standard
& 5.81 & 5.04 & {14.26} &  22.39 & 3 \\

\bottomrule
\end{tabular}
\end{table}

\subsection{MoE Projection Layer}

\begin{itemize}
    \item \textbf{Number of Experts:} 
    We evaluate \(M \in \{1, 4, 8\}\) to study how many experts are needed to represent features from different regions under a shared projection.

    \item \textbf{Concatenation:} We replace the residual addition from Eq.~(\ref{eq:moe_residual_fusion_vec}) by concatenation:  
\begin{align}
\mathbf{z}_{r_{\mathrm{in}}}
&=\mathrm{Swish}\!\left(\mathrm{Concat}\left(
\hat{\mathbf{z}}^{\mathrm{price}}_{r_{\mathrm{in}}},
\hat{\mathbf{z}}^{\mathrm{exo}}_{r_{\mathrm{in}}}\right)\right)
\in \mathbb{R}^{2h}.
\label{eq:concat_eq1}
\end{align}
    \item \textbf{Cross-Attention:} We apply multi-head attention with 
    \({\mathbf{X}}^{\mathrm{price}}_{r_{\mathrm{in}}}\) as the query and 
    \({\mathbf{X}}^{\mathrm{exo}}_{r_{\mathrm{in}}}\) as both key and value to produce the attention fused feature:
    \begin{equation}
        \mathbf{z}_{r_\mathrm{in}} = \mathrm{CrossAttention}\left(
            {\mathbf{X}}^{\mathrm{price}}_{r_{\mathrm{in}}},
            {\mathbf{X}}^{\mathrm{attn}}_{r_{\mathrm{in}}}
        \right).
    \label{eq:xattn}
    \end{equation}
\end{itemize}

The results in Table~\ref{table_ablation} show that using a single expert yields \textbf{6.0\%} higher AQL than using {\(M = 4\)}. Further increasing the number of experts to \(M = 8\) does not further improve the loss. This indicates that \(M = 4\) is sufficient to differentiate regional patterns. Replacing the residual addition with concatenation leads to \textbf{5.3\%} higher AQL and switching to cross-attention yields comparable performance to residual addition, while introducing additional parameters. 
This suggests that the residual addition strikes a favorable balance between predictive performance and model simplicity.

\subsection{Topology-Guided Sparse Graph Mask}

\begin{itemize}
    \item \textbf{Random Graph Mask:} We replace Eq.~(\ref{eq:sparse_mask_vector}) with a randomly sampled vector, where each decay weight is drawn independently from a uniform distribution over $[0, 1]$, thereby removing the spatial graph prior:
    \begin{equation}
        \mathbf{m}_{r_{\mathrm{out}}} \sim \mathcal{U}(0, 1)^{|\mathcal{R}| \times 1}.
    \label{eq:random_decay_mask}
    \end{equation}

    \item \textbf{No Graph Mask:} We remove the decay mask, which simplifies Eq.~(\ref{eq:sparse_avg_vectorized}) to a uniform average over input regions:
    \begin{equation}
        \mathbf{U}_{r_{\mathrm{out}}}
        =
        \frac{
        \mathbf{1}^\top \mathbf{S}
        }{
        |\mathcal{R}|
        },
    \label{eq:uniform_avg}
    \end{equation}
\end{itemize}

The results in Table~\ref{table_ablation} demonstrate that randomizing or removing the graph decay mask leads to a significant drop in all metrics.
We also observe that such results are on par with those of GNN baselines.
This suggests that relying on purely data-driven spatial mixing without an explicit graph-based constraint may introduce noisy signals from weakly related regions, increasing the risk of overfitting and ultimately limiting the model's performance.

\subsection{Hierarchical Quantile Head}

\begin{itemize}
    \item \textbf{Non-Negative Functions:}  
    We replace the absolute-value function used in Eq.~(\ref{eq:upper_add}) and (\ref{eq:lower_add}) with ReLU.

    \item \textbf{Standard Multi-Quantile Head:}  
    The Eq.~(\ref{eq:res_upper}) and (\ref{eq:res_lower}) are skipped, and \(\mathbf{U}_{r_{\mathrm{out}}}\) is passed directly to independent dense layers to produce quantile trajectories.

\end{itemize}

The results in Table \ref{table_ablation} reveal that replacing the absolute-value function with ReLU does not result in a noticeable change in overall performance, suggesting that the choice of non-negative function is flexible. Moreover, while the hierarchical quantile head achieves comparable loss to the standard multi-quantile head, the latter exhibits a mean AQCR of \textbf{5.04\%}, indicating that the hierarchical design mitigates quantile crossing without harming performance.

\section{Conclusion}
\label{sec:conclusion}
In this paper, we introduced a comprehensive, large, and up-to-date \textcolor{black}{renewable-aware} European electricity market dataset, which will benefit both the research community and the energy industry.
Furthermore, we proposed PriceFM, a foundation model pretrained on this diverse dataset, showing better generalizability against multiple competitive baselines. Extensive experiments and ablation studies highlight the importance of spatial context and individual contribution of design choices. 
\textcolor{black}{In this context, PriceFM provides a scalable probabilistic forecasting tool for power systems with increasing renewable penetration. By jointly modeling solar, wind, load, electricity prices, and transmission-topology-induced regional interactions, the proposed framework supports uncertainty-aware decision-making for renewable integration and flexible grid operation.}

Looking ahead, if the physical transmission network evolves over time (every few years), the proposed framework can be adapted by only updating the adjacency matrix and retraining the model, while the overall pipeline remains unchanged.
In addition, alternative notions of spatial relatedness may further improve performance.
For example, one could design weighting schemes based on socio-economic or system characteristics, such as population or power flow, potentially yielding better cross-region aggregation.

\section*{Appendix}

\subsection{Hardware and Computation}
\label{appendix:hardware}
The PriceFM is evaluated on both an \textbf{NVIDIA A100 GPU} and an \textbf{Intel Core i7-1265U CPU}, respectively. The NVIDIA A100 is designed for high-performance computing and deep learning workloads, offering 80~GB of high-bandwidth memory and up to 6,912 CUDA cores. In contrast, the Intel i7-1265U is a power-efficient CPU commonly found in standard laptops.  
The inference time for both setups is under \textbf{10 seconds}. However, we note that the computation time is not critical for our application, as bid submissions can occur at any point before the market gate closure on a daily basis.

\subsection{Adjacency Matrix}
\label{spatial_adj}

We model the European market as a graph \( G=(\mathcal{R}, \mathcal{E}) \), where each node \( r \in \mathcal{R} \) is a bidding zone and edges indicate direct power flow via cross-border interconnections. This spatial topology is detailed in Table \ref{tab:neighbors}. Let \( \mathcal{N}(r) \) denote the set of directly connected neighbors of \( r \), excluding \( r \) itself. 
The binary adjacency matrix \( A \in \{0,1\}^{|\mathcal{R}|\times|\mathcal{R}|} \) is defined by
\begin{equation}
A_{r,s} =
\begin{cases}
1, & \text{if } s \in \mathcal{N}(r),\\
0, & \text{otherwise,}
\end{cases}
\quad r,s \in \mathcal{R}.
\end{equation}
For GNN layers, self-loops can be added via \( \tilde{A} = A + I \).

\begin{table}[t]
\begin{center}
\caption{Direct neighbors by region.}
\label{tab:neighbors}
\begin{tabular}{ll}
\toprule
\textbf{Code} & \textbf{Direct Neighbors} \\ 
\midrule
AT      & CZ, DE-LU, HU, IT-NORD, SI \\
BE      & DE-LU, FR, NL \\
BG      & GR, RO \\
CZ      & AT, DE-LU, PL, SK \\
DE-LU   & AT, BE, CZ, DK1, DK2, FR, NL, NO2, PL, SE4 \\
DK1     & DE-LU, DK2, NL, NO2, SE3 \\
DK2     & DE-LU, DK1, SE4 \\
EE      & FI, LV \\
ES      & FR, PT \\
FI      & EE, NO4, SE1, SE3 \\
FR      & BE, DE-LU, ES, IT-NORD \\
GR      & BG, IT-SUD \\
HR      & HU, SI \\
HU      & AT, HR, RO, SI, SK \\
IT-CALA & IT-SICI, IT-SUD \\
IT-CNOR & IT-CSUD, IT-NORD \\
IT-CSUD & IT-CNOR, IT-SARD, IT-SUD \\
IT-NORD & AT, FR, IT-CNOR, SI \\
IT-SARD & IT-CSUD \\
IT-SICI & IT-CALA \\
IT-SUD  & GR, IT-CALA, IT-CSUD \\
LT      & LV, PL, SE4 \\
LV      & EE, LT \\
NL      & BE, DK1, DE-LU, NO2 \\
NO1     & NO2, NO3, NO5, SE3 \\
NO2     & DE-LU, DK1, NL, NO1, NO5 \\
NO3     & NO1, NO4, NO5, SE2 \\
NO4     & FI, NO3, SE1, SE2 \\
NO5     & NO1, NO2, NO3 \\
PL      & CZ, DE-LU, LT, SE4, SK \\
PT      & ES \\
RO      & BG, HU \\
SE1     & FI, NO4, SE2 \\
SE2     & NO3, NO4, SE1, SE3 \\
SE3     & DK1, FI, NO1, SE2, SE4 \\
SE4     & DE-LU, DK2, LT, PL, SE3 \\
SI      & AT, HR, HU, IT-NORD \\
SK      & CZ, HU, PL \\
\bottomrule
\end{tabular}
\end{center}
\end{table}

\subsection{Hyperparameter Optimization}
\label{appendix:Hyperparameters}
All models are optimized based on validation loss, and the checkpoint with the lowest validation loss is saved. We use the Adam optimizer with a default learning rate of \(1 \times 10^{-3}\). Models are trained for 20 epochs with a batch size of 128. We empirically vary the learning rate to \(1 \times 10^{-4}\) and \(4 \times 10^{-3}\), and the batch size to 32 and 64, and observe that for batch sizes $\leq 128$, the lowest validation loss across all models can consistently be reached within 20 epochs. 
Moreover, smaller batch sizes typically converge in fewer epochs, but require longer training time. 
Therefore, we recommend setting the number of epochs to 20 and the batch size to 128 as a good trade-off.
The search space of other hyperparameters varies by model and is summarized in Table~\ref{tab:hyperparams_all}.

\begin{table}[ht]
\begin{center}
\caption{Hyperparameter search space.}
\label{tab:hyperparams_all}
\begin{tabular}{ll}
\toprule
\textbf{Model} & \textbf{Search Space} \\
\midrule
PriceFM &
\begin{tabular}[t]{@{}l@{}}
hidden\_size: \{24, 72, 168\} \\
n\_layers: \{2, 3, 4\} \\
n\_experts: \{1, 4, 8\} \\
graph\_degree\_cutoff: \{0, 1, $\ldots$, 10\}
\end{tabular}
\\
\midrule

FEDFormer &
\begin{tabular}[t]{@{}l@{}}
hidden\_size: \{32, 128, 512\} \\
conv\_hidden\_size: \{32, 128, 512\} \\
e\_layers: \{2, 3, 4\} \\
n\_heads: \{2, 4, 8\} \\
dropout: \{0.1, 0.3, 0.5\}
\end{tabular}
\\
\midrule

iTransformer &
\begin{tabular}[t]{@{}l@{}}
hidden\_size: \{32, 128, 512\} \\
e\_layers: \{2, 3, 4\} \\
d\_ff: \{512, 1024, 2048\} \\
n\_heads: \{2, 4, 8\} \\
dropout: \{0.1, 0.3, 0.5\}
\end{tabular}
\\
\midrule

PatchTST &
\begin{tabular}[t]{@{}l@{}}
hidden\_size: \{32, 128, 512\} \\
e\_layers: \{2, 3, 4\} \\
n\_heads: \{2, 4, 8\} \\
dropout: \{0.1, 0.3, 0.5\} \\
patch\_len: \{4, 6, 12\}
\end{tabular}
\\
\midrule

TimesNet &
\begin{tabular}[t]{@{}l@{}}
hidden\_size: \{32, 128, 512\} \\
conv\_hidden\_size: \{32, 128, 512\} \\
e\_layers: \{2, 3, 4\} \\
dropout: \{0.1, 0.3, 0.5\}
\end{tabular}
\\
\midrule

TimeXer &
\begin{tabular}[t]{@{}l@{}}
hidden\_size: \{32, 128, 512\} \\
e\_layers: \{2, 3, 4\} \\
n\_heads: \{2, 4, 8\} \\
d\_ff: \{512, 1024, 2048\} \\
dropout: \{0.1, 0.3, 0.5\}
\end{tabular}
\\
\midrule

GraphConv &
\begin{tabular}[t]{@{}l@{}}
hidden\_size: \{32, 128, 512\} \\
layers: \{2, 3, 4\} \\
dropout: \{0.1, 0.3, 0.5\}
\end{tabular}
\\
\midrule

GraphAttn &
\begin{tabular}[t]{@{}l@{}}
hidden\_size: \{32, 128, 512\} \\
layers: \{2, 3, 4\} \\
n\_heads: \{2, 4, 8\} \\
dropout: \{0.1, 0.3, 0.5\}
\end{tabular}
\\
\midrule

GraphSAGE &
\begin{tabular}[t]{@{}l@{}}
hidden\_size: \{32, 128, 512\} \\
layers: \{2, 3, 4\} \\
aggregate: \{mean, max, sum\}
\end{tabular}
\\
\midrule

GraphDiff &
\begin{tabular}[t]{@{}l@{}}
diff\_steps: \{2, 4, 6\} \\
hidden\_size: \{32, 128, 512\} \\
layers: \{2, 3, 4\}
\end{tabular}
\\
\midrule

GraphARMA &
\begin{tabular}[t]{@{}l@{}}
hidden\_size: \{32, 128, 512\} \\
layers: \{2, 3, 4\} \\
order: \{1, 2, 4\} \\
iteration: \{1, 2, 4\}
\end{tabular}
\\

\bottomrule
\end{tabular}
\end{center}
\end{table}

\printbibliography[heading=bibintoc,title=Reference]

\end{document}